\documentclass{PoS}
\usepackage{latexsym}
\usepackage{amsmath}
\usepackage{epsfig,graphics}

\newcommand{\be}{\begin{equation}}
\newcommand{\ee}{\end{equation}}

\title{Large N}

\ShortTitle{Large N}

\author{\speaker{Michael Teper}\\
        Dept of Physics, University of Oxford, OX1 3NP, UK\\
        E-mail: \email{m.teper1@physics.ox.ac.uk}}

\abstract{I review some of the things we have learned about
large $N$ gauge theories (and QCD$_\infty$) from lattice calculations in
recent years. I point to some open problems.}

\FullConference{The XXVI International Symposium on Lattice Field Theory \\
                July 14 - 19, 2008\\ 
                Williamsburg, Virginia, USA}

\begin{document}

\section{Introduction}

The fact that QCD has no obvious expansion parameter,
(the coupling sets the length scale and cannot otherwise
be independently varied) led 't Hooft to suggest 
\cite{tHooft}
a parameter that is not obvious: $1/N$ where $N$ is the number 
of colours. One thinks of the SU($N$) theory as being
the same as the SU($\infty$) theory up to  corrections 
that are $O(1/N^2)$, or $O(1/N)$ in $QCD_N$. There
are a number of ill-understood features of the strong
interactions, such as the OZI rule, that become
obvious and exact at $N=\infty$ 
\cite{largeN}.
However all these `successes' require that $N=3$ should be 
`close to' $N=\infty$; something that is not at all guaranteed.
In particular, while the $N=\infty$ theory is simpler,
it is unfortunately not simple enough to have been solved.
There are however a number of things one can say on
simple counting arguments 
\cite{tHooft,largeN}.
For example analysing perturbation theory to all orders
tells us that to achieve a smooth $N=\infty$ limit one should
keep the 't Hooft coupling, $\lambda = g^2N$, fixed as  
$N\to\infty$. Simple combinatorics (the colour singlet
combination becomes vanishingly unlikely as $N\to\infty$) 
tells us that in the confining phase of that limit mesons 
and glueballs do not mix or decay, and indeed there are no 
colour singlet interactions at all. The lack of scattering makes it
conceivable that integrability might play a role at $N=\infty$.
Of course, all this depends on the  $N=\infty$ theory having 
a confining phase -- again something that is not guaranteed.

During the past decade, the large $N$ limit has been at the center
of one of the most exciting theoretical developments: the
strong-weak coupling duality of AdS/CFT and its derivatives. (See
\cite{adscft}
for a review.)
As attempts are made to develop supergravity duals of
non-conformal large-$N$ field theories
it is important to know something about the detailed properties 
of those theories to test the success of these attempts.

There are some simple questions here that lattice methods can help 
to answer. In this brief review I will say something about 
the following with some stress on open, accessible problems. \\
$\bullet$ Is large-$N$ confining?\\
$\bullet$ Is SU(3) really close to SU($\infty$)?\\
$\bullet$ Mesons and QCD as $N\to\infty$.\\
$\bullet$ Space-time reduction at large-$N$.\\
$\bullet$ Hot SU($N$) gauge theory. \\
$\bullet$ Interlacing $\theta$-vacua.\\
$\bullet$ The spectrum of closed flux tubes in $D=3,4$.

The list of things I will not discuss is (unfortunately!)
much longer. For example I will not discuss the work
on large-$N$ phase transitions that has been reviewed in
Lattice 2005 and in Lattice 2007
\cite{Lat0507N}.
Neither will I discuss numerical tests
\cite{BBMT_KN}
of the remarkably successful analytic calculations
by Karabali-Nair of string tensions in $D=2+1$
\cite{KN}.
Nor will I review $k$-string tensions
\cite{kstring,blmtuwG},
in $D=2+1$ or $D=3+1$, nor high-$T$ domain wall tensions
\cite{sigmadd}; 
nor chiral symmetry breaking
\cite{NN_chi}, 
and the role of topology
\cite{ncmtuw_chi}, 
nor $\cdots$.

Throughout my talk I will try to point to problems which 
are accessible, interesting and are awaiting your involvement! 

\section{Is $N=\infty$ confining? Is $N=3$ close to $N=\infty$? }

At $N=\infty$ decay widths are zero and we have a prefect OZI rule.
This might explain why in QCD we can have modest decay widths,
e.g. $\Gamma_\rho/m_\rho \sim 1/5 \ll 1$, and a good OZI rule.
At large $N$ the explanation is essentially combinatoric: if
the theory is confining, so that all states are colour singlets, 
we need the $q\bar{q}$ that pops out of the vacuum to have the 
right colour to make two colour singlet $\pi$ when combined with 
the $q,\bar{q}$ in the $\rho$, and there is clearly a $1/N$ `phase
space' suppresssion there. If the theory is not confining any
bound state can decay without any suppression into coloured states.

Is SU($\infty$) confining? A pedestrian but reliable approach to
this question is to repeat for SU(4), SU(5), ...  the kind of 
calculations that answered this question for SU(2) and SU(3). Let 
me show you an example for SU(6)
\cite{hmmt}.

We are on a $D=3+1$ hypertorus. Suppose its spatial dimensions are 
$l\times l^2_\perp $. Consider one unit of fundamental flux
wrapped around the $l$-torus. If the theory is linearly confining 
this becomes a flux tube of length $l$ and we expect its ground state 
energy to vary with $l$ as
\begin{equation}
E_0(l)
=
\sigma l -  c_{eff}(l)\frac{\pi}{6}\frac{D-2}{l}
\stackrel{l\to\infty}{=} 
\sigma l -  \frac{\pi}{6}\frac{D-2}{l}
\label{eqn_luscher} 
\end{equation} 
where the correction term arises from the zero-point energies
of the massless transverse oscillations (the Luscher correction)
and we have assumed that the only massless modes along the flux tube
are those of these stringy oscillations. This is the universality 
class of the simple free bosonic string, and it is important to
determine numerically whether in fact $C_{eff}(l)\stackrel{l\to\infty}{=}1$.

\begin	{figure}[h!]
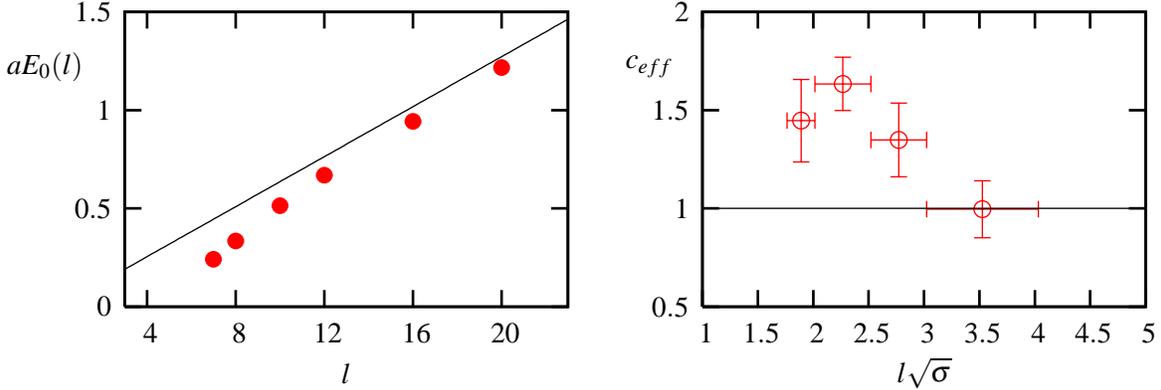

\begin	{center}
\leavevmode
\centerline{{\input {plot_K1b} } \hspace{-15mm}  {\input {plot_Ceffn6} }}
\end	{center}
\vskip -0.5cm
\caption{SU(6) : flux loop energy (left); effective Luscher correction (right).}
\label{fig_K1n6d4}
\end 	{figure}

In Fig.~\ref{fig_K1n6d4} I show you what one obtains for SU(6).
\cite{hmmt}.
The approach to linearity at large $l$ is evident. And we see that
the deviations from linearity at smaller $l$, can be described
by a $c_{eff}(l)$ that appears to approach unity at larger $l$. So
it appears that the $N=\infty$ gauge theory is linearly confining
and is in the bosonic string universality class.

\begin	{figure}[h!]
\begin	{center}
\leavevmode
\input	{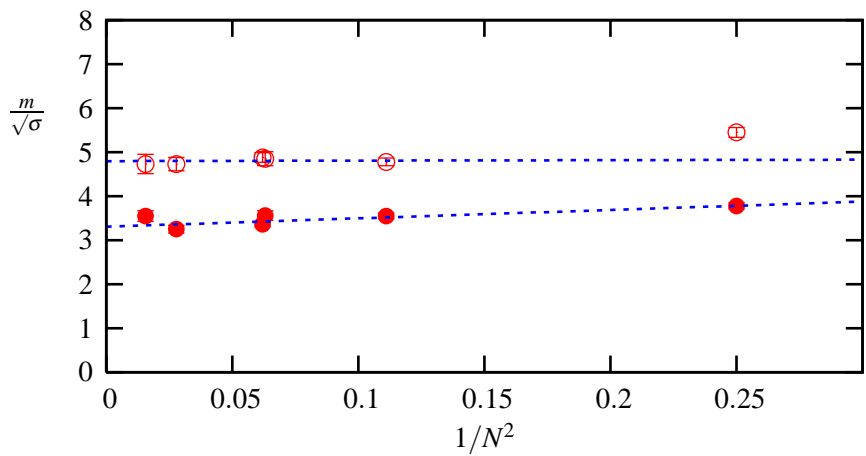}
\end	{center}
\vskip -0.5cm
\caption{Continuum glueball masses in units of the string tension
versus $1/N^2$.}
\label{fig_gkNd4}
\end 	{figure}

This is a good start. However for large-$N$ to be phenomenologically
relevant, we need to show that for typical physical quantities
the difference between SU(3) and SU($\infty$) is `small'. So let
us calculate the lightest $J^{PC}=0^{++},2^{++}$ glueball masses,
express them in units of the simultaneously calculated string tension,
and extrapolate the ratios to the continuum limit so as to obtain
values of  $m/\surd\sigma$. Now repeat this for various $N$. The
leading large-$N$ correction should be $O(1/N^2)$, so plot
the resulting ratios 
\cite{blmtG,blmtuwG}
against $1/N^2$ in Fig.~\ref{fig_gkNd4}. (For a very detailed 
comparison of SU(3) and SU(8) glueball spectra see
\cite{meyer}.)
We observe in  Fig.~\ref{fig_gkNd4} that the $O(1/N^2)$ corrections 
are indeed small, SU(3) appears to be `close to' SU($\infty$), 
so large-$N$ does appear to be physically relevant.
 
\section{QCD :  $N=\infty$}

Because quarks are in the fundamental, some corrections to QCD
are $O(1/N)$ rather than $O(1/N^2)$, i.e. the size of the corrections 
might be closer to those of SU(2) than SU(3) in Fig.~\ref{fig_gkNd4}.
Even so, we see from  Fig.~\ref{fig_gkNd4} that this is modest.

Since QCD$_{\infty}$ is a (unitary) quenched theory, we can approach
it through a sequence of quenched calculations of increasing $N$.
These will generally have $O(1/N^2)$ corrections. If one calculates
hadron masses at various $N$ and extrapolates to $N=\infty$
at various fixed values of $m_q$, one can then do the usual chiral
extrapolation in that limit. 

A first step might be to do calculations not in the continuum 
limit but at some fixed small value of $a \surd\sigma$. 
And, if the calculation is not trying to be too precise,
one can do chiral extrapolations at fixed $N$ without worrying
about the subtleties. In this way one gets the hadron spectrum at 
$N=\infty$, in units of $\surd\sigma$ or $r_0$. One can then 
compare it either to experiment or to full QCD lattice calculations, 
to see how large are the $O(1/N)$ corrections.

\begin	{figure}[htb]
\begin	{center}
\leavevmode
\centerline{
\includegraphics[scale=.7,angle=360,width=7cm]{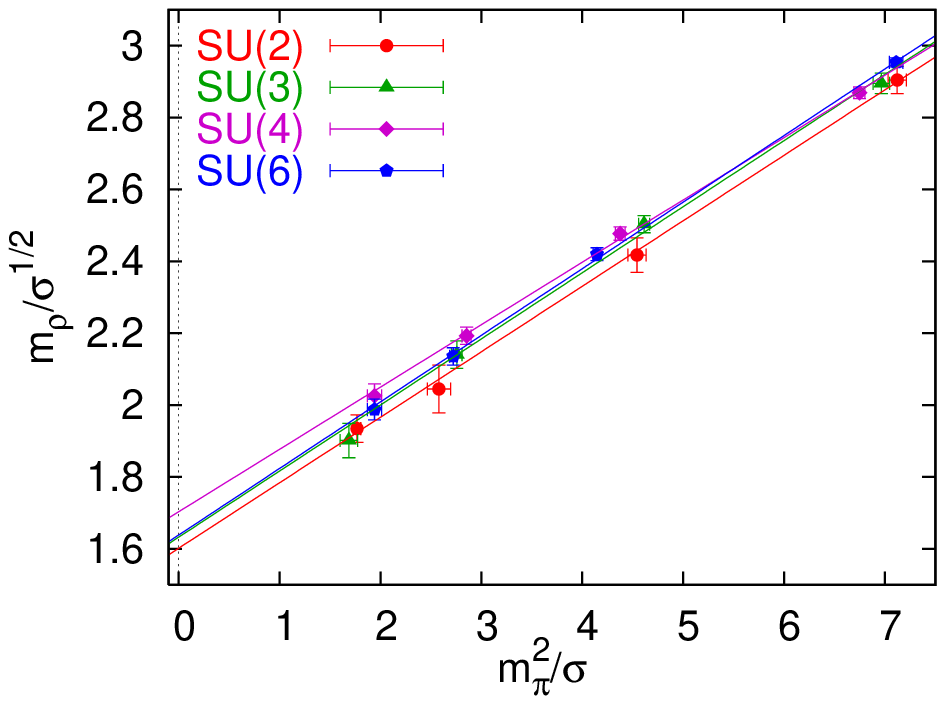} \hspace{5mm} \includegraphics[scale=.7,angle=360,width=7cm]{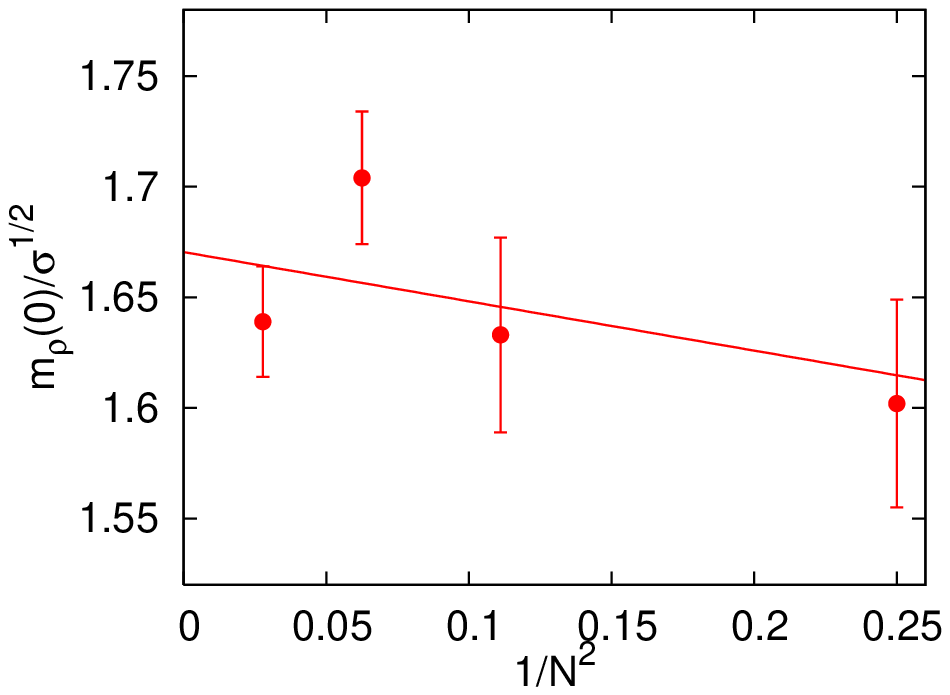} }
\end	{center}
\caption{$m_\rho$ versus $m_\pi$ (left); $m_\rho$ for $m_q=0$ versus $1/N^2$ (right). }
\label{fig_qcd}
\end 	{figure}

There have recently been two pioneering calculations of the latter kind 
\cite{qcd_ldd,qcd_fbgb}
that calculate $m_\rho/\surd\sigma$. (Also $m_\pi$, but that is traded
for the physical $m_q$.)
In Fig.~\ref{fig_qcd} I show some figures from
\cite{qcd_fbgb}.
On the left is a chiral plot of $m_\rho$ versus $m^2_\pi$ for $N\in [2,6]$
and on the right is the large $N$ extrapolation of $m_\rho/\surd\sigma$.
The $N$-dependence is clearly very weak, but this is expected for
the quenched theory. More to the point, the $N=\infty$ chiral value is
\begin{equation}
\frac{m_\rho}{\surd\sigma} 
=
1.670(24)
\sim
735 {\mathrm{MeV}}
\qquad;\qquad \surd\sigma \simeq 440 {\mathrm{MeV}}
\label{eqn_mrho} 
\end{equation} 
which is within $\sim 35 {\mathrm{MeV}} \sim \Gamma_\rho/4$
of the experimental value. These calculations 
\cite{qcd_ldd,qcd_fbgb}
thus provide explicit evidence that, as long hoped, QCD is close 
to QCD$_\infty$.

Let me list a few of the many interesting questions that larger scale
$N\to\infty$ calculations of this kind could address:\\
$\bullet$ Scalar mesons as $N\to\infty$ : do the $\leq 1 {\mathrm{GeV}}$
states disappear? \\
$\bullet$ The scalar nonet and the place of lightest scalar glueball? \\
$\bullet$ Flavour singlet tensor and pseudoscalar mesons and glueballs? \\
$\bullet$ Excited states stable $\longrightarrow$ Regge trajectories? \\
$\bullet$ Excited states stable $\longrightarrow$ clean meson 
excitation spectrum. \\
$\bullet$ SU($2n_f$) baryon (Dashen-Manohar) symmetry as $N\to 3$. \\
(This last might require full $QCD_N$ -- the next step.) Note that the
suppression of decays and mixings as $N\uparrow$ might clarify
many questions about the hadron spectrum that are obscured by
large decay widths and mixings, and make such lattice calculations
much cleaner and simpler at larger $N$ than at $N=3$. 
Answering the above questions would help answer some long-standing 
questions in hadron spectroscopy, e.g. are the  $\leq 1 {\mathrm{GeV}}$
scalar mesons molecular? Do the three $0^{++}$ flavour singlets  
$\in 1.3-1.7 {\mathrm{GeV}}$ arise from the scalar glueball mixing
with two scalar nonet mesons? And this would give phenomenologists 
some concrete idea of how and where to look for the lightest tensor 
and pseudoscalar glueballs. Finally, the $N=\infty$ mesonic spectrum
might show simple patterns that will provoke more theoretical
understanding of that theory.

\section{Calculating as $N\to\infty$ : how much harder?}

First some obvious counting.
The pure gauge theory is dominated by the basic operation of multiplying
two $N\times N$ matrices together, i.e. $O(N^3)$ operations.
(The Cabibbo-Marinari heat-bath update of all SU(2) subgroups 
can be done in just $O(N^2)$ operations.) Current quenched 
QCD$_N$ calculations are, by contrast, dominated by the fermionic
part where one is multiplying matrices times vectors, 
i.e. $O(N^2)$ operations.

Second, some less obvious counting.
We calculate masses from connected correlators and in the pure 
gauge theory these are $O(1/N^2)$. Does this mean we need to
increase the statistics to maintain the signal as $N\uparrow$?
The answer is no: the statistical fluctuations are themselves 
related to higher order correlation functions and one can show 
\cite{blmtuwG}
that they have the same $O(1/N^2)$ dependence: the error/signal
ratio is independent of $N$. Indeed if one compares actual
calculations of masses at different $N$ one finds
\cite{blmtuwG}
very little $N$-variation in the error/signal ratio when one
goes from SU(2) all the way to SU(8). One can apply the
same analysis to meson propagators in quenched QCD. Just as
the $t$-dependence of the error/signal ratio is very different 
for glueball and meson propagators (it grows exponentially for 
the former but is nearly constant for the latter) one
finds a different $N$-dependence: while it is $O(N^0)$ for
the former one can show
\cite{qcd_fbgb}
that it is $O(1/N)$ for the latter. In practice the observed decrease
\cite{qcd_fbgb}
is closer to  $O(1/\sqrt{N})$. So for $QCD_N$ there is a gain.

In summary: the cost of pure gauge calculations is $\propto N^3$ while
that of (quenched) QCD is $\propto N^{1.5}$ or even $\propto N$,
unless one goes to very large $N$, where it will revert to $N^3$.

There are other less quantifiable gains at larger $N$. The
existence of a first order strong-weak coupling transition
for $N\geq 5$ means that one has some confidence about where one
can start applying weak-coupling (Symanzik-type) analyses. 
The rapid disappearance of instantons with $\rho \sim O(a)$
as $N\uparrow$ 
\cite{ncmtuw}
means that exceptional configurations (with their accompanying
`exceptional' fixes) should rapidly disappear. Also, the fact that
finite volume corrections disappear as $N\to\infty$ means
that we can work on smaller volumes. And the fact that excited states 
become stable means that they will be much easier to identify.

Pure gauge lattice calculations, with continuum limits and large $N$
extrapolations of these, can and have been carried out on
a few PC's. The quenched QCD calculations have used small
clusters (or equivalents).

\section{$\lambda = g^2N$ fixed as $N\to\infty$?}

\begin	{figure}[h!]
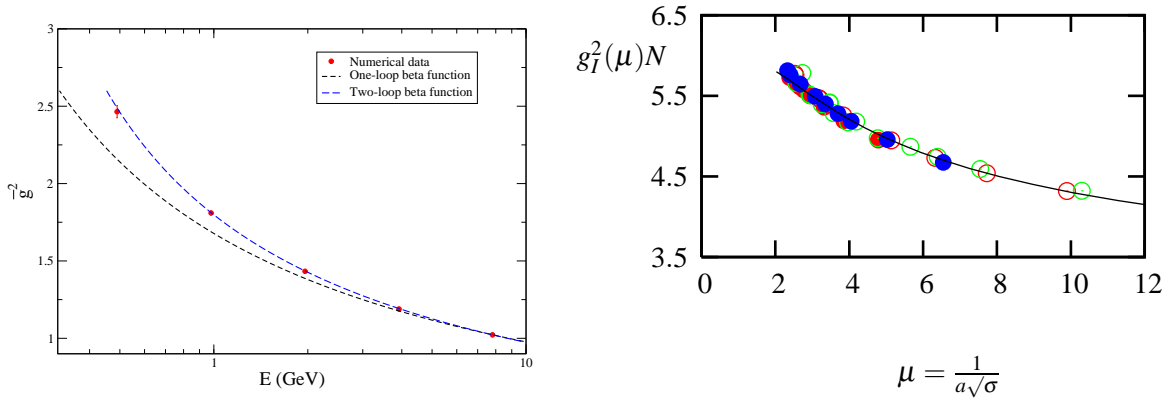

\begin	{center}
\leavevmode
\centerline{\scalebox{0.70}{\epsfig{figure=coupling.eps, angle=360, width=10cm} } \hspace{-5mm}  {\input	{plot_gIN}}}
\end	{center}
\vskip -0.5cm
\caption{SF running coupling for SU(4) (left); bare 't Hooft running coupling
for $N\in [2,8]$ (right).}
\label{fig_gN}
\end 	{figure}

Counting diagrams tells us that to have a smooth $N\to\infty$
limit we should keep $g^2N$ constant. For a running coupling 
that means keeping $g^2N(\mu)$ constant at constant
$\mu/\surd\sigma$. That is to say, if we plot $g^2N(\mu)$
against $\mu/\surd\sigma$ the result should not depend on $N$ up
to $1/N^2$ corrections. In Fig~\ref{fig_gN} we show how the 
(mean-field improved) bare coupling $g^2_I N$ (a running coupling 
on the length scale $a$) varies with the scale, and it 
displays exactly such an $N$-independence
\cite{blmtG}.
In
\cite{camtat}
it has been shown how the usual $\beta$-function with modest lattice
corrections describes the running in  Fig~\ref{fig_gN} and this allows 
the calculation of $\Lambda_{\overline{MS}}$ for all $N$.

\vskip 0.25cm

Of course it would be nice to have a continuum $\beta$-function
calculation at larger $N$. That has recently been provided in
\cite{blgm}
for SU(4) using the Schrodinger functional definition. I reproduce
it in Fig~\ref{fig_gN}. Coupled with earlier SU(2) and SU(3) results,
this leads to
\cite{blgm}
\begin{equation}
\frac{\Lambda_{\overline{MS}}}{\surd\sigma} 
=
0.528(40) + \frac{0.18(36)}{N^2}
\label{eqn_lamN} 
\end{equation} 
which is entirely consistent with the values obtained in
\cite{camtat}.

Question: does the $N=\infty$ SF coupling acquire non-perturbative jumps
at the Narayanan-Neuberger
\cite{Lat0507N}
finite volume phase transitions?

\section{Hot gauge theories at large-$N$}

Calculations 
\cite{blmtuwT}
for $N\in [2,8]$ show that the deconfining transition
is first order for $N\geq 3$ with a latent heat of $O(1)$
when expressed in natural units. Now, at $T_c$ the free energies
of the gluon plasma and the confined ensembles are equal:
$F_{gp}\stackrel{T=T_c}{=}F_{conf}$. So, since we expect 
$F_{gp}\propto N^2$ we are only interested in the $\propto N^2$ 
piece of $F_{conf}$ at $T_c$ when $N\to\infty$ . The only piece 
in the confined phase that has this dependence on $N$ 
is the vacuum energy density, which
provides one conventional definition of the gluon condensate. 
(Without this we could expect $T_c\to 0$ as $N\to \infty$.)
So if the gluon plasma was weakly coupled, we could use the 
usual Stefan-Boltzmann (SB) expression for $F_{gp}$ and thus obtain
$T_c$ in terms of the vacuum energy density and the gluon
condensate -- which would be a very nice prediction! Unfortunately 
for us (fortunately for our AdS/CFT colleagues - see below) the
plasma turns out to be strongly coupled near $T_c$. 

\begin	{figure}[h!]
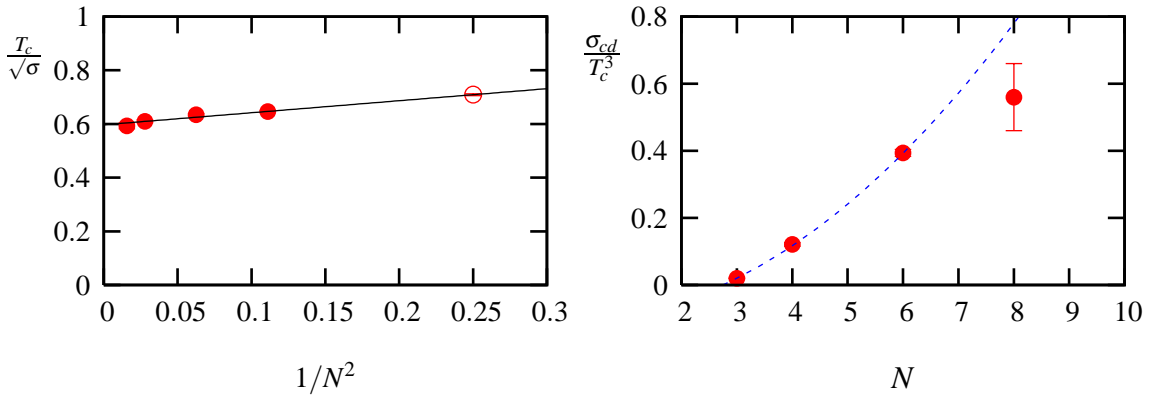

\begin	{center}
\leavevmode
\centerline{{\input {plot_tcN} } \hspace{-15mm}  {\input {plot_sigmaN} }}
\end	{center}
\vskip -0.5cm
\caption{$N$-dependence of the deconfining temperature (left) and
the interface tension (right).}
\label{fig_tNd4}
\end 	{figure}

The deconfining temperature is shown as a function of $1/N^2$
in Fig.~\ref{fig_tNd4}
\cite{blmtuwT}.
We see modest corrections to the $N=\infty$ 
limit: the fit is $T_c/\surd\sigma = 0.597(4) + 0.45(3)/N^2$.
Remarkably this even fits SU(2), where the transition is second order. 

We also show in  Fig.~\ref{fig_tNd4} the interface tension
\cite{blmtuwT}
between the confined and deconfined phases. We see that the SU(3)
value is strongly suppressed: indeed the fit shown is
$\sigma_{cd}/T^3_c = 0.0138N^2 - 0.104\stackrel{N=3}{\simeq} 0.020$.
This suppression is much stronger than the modest suppression of the
SU(3) latent heat, and is presumably the real reason why that
transition has looked so `weakly' first order in the past.

One finds a very similar pattern of results in $D=2+1$
\cite{liddle_holland}
except that now it is SU(4) that is weakly first order and, as 
expected, $T_c$ is larger, $T_c/\surd\sigma = 0.903(3) + 0.88(5)/N^2$.

For quite a large range of $T$ above $T_c$ we have a non-trivial 
strongly coupled (quark-)gluon plasma, that is being probed by 
experiments at RHIC and (soon) LHC. Calculating the kinetic
properties of this plasma has become the area of choice for 
gauge-gravity applications
\cite{dsas}:
we have a strongly coupled system, and
finite $T$ breaks the supersymmetry (fermions become massive, 
scalars are then unprotected and also become massive), 
so the resulting theory (possiby rescaling the number of degrees 
of freedom) may be `not so different' from QCD. However one major
issue is that gauge-gravity dualities tell us about $N=\infty$
not $N=3$. So, is the observed strongly coupled plasma a feature
of just SU(3), or does it survive as $N\to\infty$?

\begin	{figure}[htb]
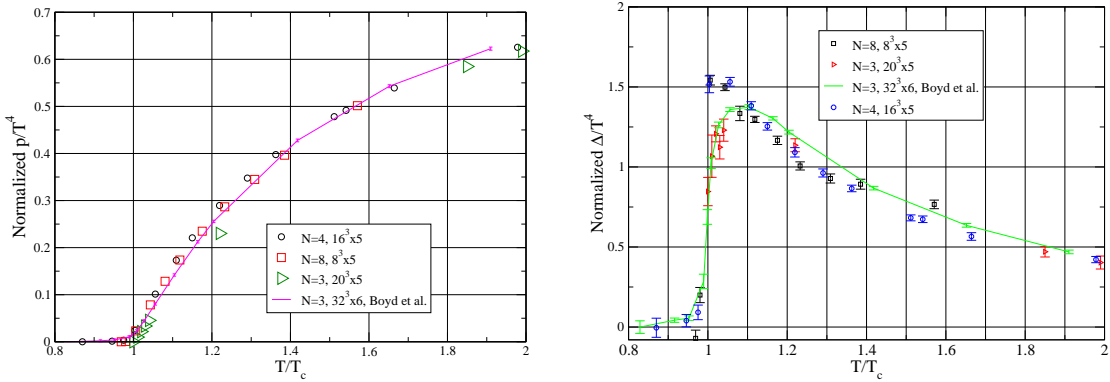

\begin	{center}
\leavevmode
\centerline{
\includegraphics[scale=.7,angle=360,width=7cm]{P348} \hspace{5mm} \includegraphics[scale=.7,angle=360,width=7cm]{Delta348} }
\end	{center}
\vskip -0.5cm
\caption{$N$-dependence of pressure (left), $\epsilon - 3p$ (right) for $T>T_c$.}
\label{fig_pN}
\end 	{figure}

This answer is that it survives as $N\to\infty$
\cite{bbmtTN}.
In Fig.~\ref{fig_pN} we see how the (lattice SB normalised) pressure
varies with $N$
\cite{bbmtTN};
clearly $SU(3) \simeq SU(\infty)$ for this quantity that provides
one characteristic measure of strong coupling. For balance I
also show $\Delta = \epsilon - 3p$ which is a measure of the
breaking of conformality: that also suvives large $N$. Nonetheless
this provides an important piece of the motivation for believing
AdS/CFT may be applicable to this physics. 

The fact that the pressure `anomaly' survives at $N=\infty$ has
implications for the dynamics; for example, at $N=\infty$ it cannot 
be due to instantons, or to the survival of colour singlet  hadrons.

\vspace*{0.5cm}

The critical result for motivating AdS/CFT for $T>T_c$ was the 
result for shear viscosity divided by entropy, $\eta/s = 1/4\pi$,
and the fact that it is close to RHIC, and is `universal' for any
theory with a gauge/gravity dual. So in both the experimental and
theoretical communities, a lattice calculation of $\eta/s$ is
eagerly awaited. As you have heard in the status report
by Harvey Meyer
\cite{hmlat08}
in his Plenary talk at this meeting, his large-scale 
lattice calculations of  $\eta/s$ look like being consistent both with 
experiment and with AdS/CFT.

It might be interesting to repeat these calculations of  $\eta/s$
for $D=2+1$ gauge theories. This might tell us whether the theory
has a gravity dual or not?

\section{Space-time reduction at large $N$}

At $N=\infty$ we have the factorisation of gauge-invariant operators: 
$\langle \mathrm{Tr}\Phi_a \mathrm{Tr}\Phi_b \rangle
=
\langle \mathrm{Tr}\Phi_a \rangle\langle\mathrm{Tr}\Phi_b \rangle$.
This implies that there is a single gauge field (gauge orbit) that 
dominates the path integral. Since physics is translation invariant,
this gauge field must also be  translation invariant (for gauge
invariant quantities). Thus one can imagine that the  $N=\infty$
theory could be defined on any volume, even infinitesimal.
On a lattice, this would be a single site: $L^4 \to 1^4$. 
This heuristic argument is made concrete in Eguchi-Kawai (EK) reduction 
\cite{ek}
which tells us that at $N=\infty$ the Schwinger-Dyson 
equations for Wilson loops on a $1^4$ lattice are the same as for the
$L=\infty$ lattice theory. This requires that the center symmetry
$U_l \to z U_l$, $z\in Z_N$, which follows because the plaquette 
is $U_\mu U_\nu U^\dagger_\mu U^\dagger_\nu$ on a $1^4$ torus,
is not spontaneously broken. To maintain this one has to impose
twisted boundary conditions (TEK)
\cite{tek}.
In the mid-1980's there were a 
number of lattice calculations using in the TEK model that
claimed to calculate physical quantities such as $T_c$ at very
large $N \sim O(64)$
\cite{das}.

Recently this has become a very active area again, and a number
of long-standing beliefs have been demolished in the process.

Firstly, it was shown in
\cite{mthv}
that at larger $N$ the center symmetry is spontaneously
broken in the TEK model, for an important range of bare couplings
that increase as $N$ increases. In
\cite{ishikawa}
it was then shown that this range appears to increase as $N\to \infty$ 
in such a way as to probably prevent a planar continuum limit. Moroever
even where the symmetry is not broken, the TEK model seems to
be stuck in the wrong phase
\cite{mthv}.
Typically the physically desired vacuum is only $O(N)$ below
undesired vacua at tree-level, and $O(N^2)$ quantum fluctuations
overwhelm this difference -- and produce insurmountable
barriers to tunnelling. So it appears that the only planar
physics you can get from TEK is the physics of an infinitesimaly
small volume 
\cite{mthv}
-- and we hardly need lattice calculations for that!

An alternative to TEK, proposed around the same time, was quenched
(in the replica sense) Eguchi-Kawai (QEK)
\cite{qek}.
Following the demise of the TEK model, there has recently been a 
renewed analysis of the QEK model in
\cite{bbss}
which has found that, in a rather more subtle way, the center symmetry 
is broken here as well. So this route to planar space-time
reduction fails as well.

That leaves two approaches. First one can deform the action so as
to bias the system towards maintaining center symetry. This
has been recently formulated in a precise way in
\cite{muly}.
This looks a promising route for reducing one or two space-time
directions, but it might not be practical for going beyond that.
It would be very interesting to see some numerical investigations
of this idea, to see what the costs are in practice (and this is
in progress
\cite{hv}).

Finally there is the `no tricks' approach to reduction
\cite{Lat0507N}.
Here you give up on complete space-time reduction, and instead
go as far as you can using the fact that finite volume corrections 
disappear as $N\to \infty$, as long as you remain on a $l^4$ volume 
that remains in the confining phase, i.e. $l > 1/T_c$. 
This means working at large enough $N$ that any $O(1/N^2)$ finite
volume corrections are smaller than your statistical errors,
i.e. very large $N$ indeed. However if you want to do that
anyway, then there is no extra cost. There has been a nice
demonstration of how this can work in a recent calculation
of the $D=2+1$ string tension, $\lim_{N\to\infty}\surd\sigma/g^2$,
in
\cite{jkrn}
performed for $N\in [21,47]$, which agrees with conventional
calculations
\cite{pap98,BBMT_KN}
extrapolating from $N\in [2,8]$. The method applies equally
well to fermions. 

Time has forced me to summarise this area as a collection of 
sound bites. However space-time reduction touches on a number
of deep and fascinating issues in field theory.

\section{Interlaced $\theta$-vacua}

Usually we act as though confining gauge theories had just one
vacuum. The reality, however, is much more interesting ...

Consider the gauge action with a $\theta$ term
\begin{equation}
S[g^2,\theta] =
\frac{1}{4g^2} \int d^4x {\mathrm{Tr}} F_{\mu\nu}F^{\mu\nu}
+
\frac{i\theta}{16\pi^2}\int d^4x 
\epsilon^{\mu\nu\rho\sigma}{ \mathrm{Tr}} F_{\mu\nu}F_{\rho\sigma}
\end{equation}
Since
\begin{equation}
\frac{1}{16\pi^2} \int d^4x 
\epsilon^{\mu\nu\rho\sigma}{ \mathrm{Tr}} F_{\mu\nu}F_{\rho\sigma}
=
Q
=
{\mathrm{integer}}
\end{equation}
we know that $\exp\{-S[\theta]\}$ and hence the vacuum
energy density $E(\theta)$ are periodic in $\theta$
\begin{equation}
E(\theta) = E(\theta + 2\pi) \qquad \forall N
\end{equation}
On the other hand, we expect that for a smooth
$N\to\infty$ limit, we need to factor $N$ from $S$ so that the
couplings to keep fixed are $1/g^2N$, $\theta/N$, ... i.e.
\begin{equation}
E(\theta) = N^2 h(\theta/N) 
\end{equation}
How do we reconcile these two apparently irreconciliable demands?

\begin	{figure}[h!]
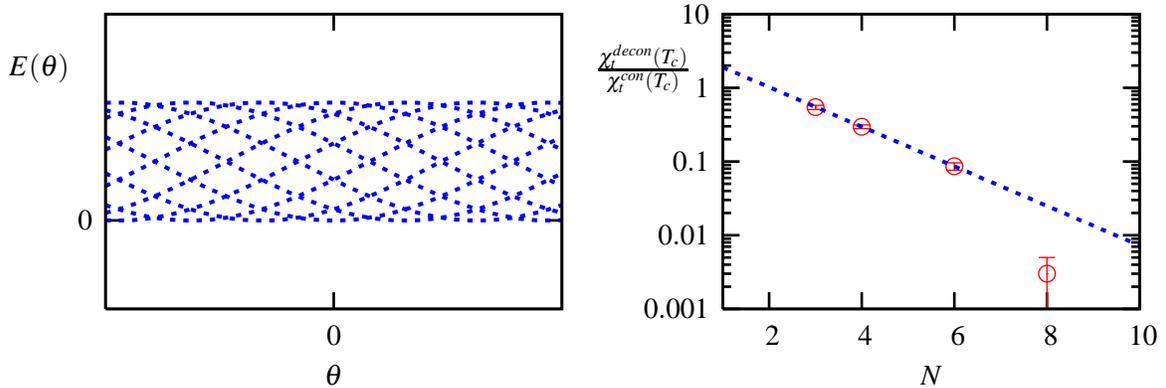

\begin	{center}
\leavevmode
\centerline{{\input {plot_multiv} } \hspace{-15mm}  {\input {plot_chiNTc} }}
\end	{center}
\vskip -0.5cm
\caption{Interlaced $\theta$-vacua for SU(10) (left); topological 
suseptibility across $T_c$ (right).}
\label{fig_multiv}
\end 	{figure}

Witten's suggestion
\cite{EW_theta}
is that $E(\theta)$ is a multi-branched function
\begin{equation}
E_k(\theta) = N^2 h\left(\frac{\theta + 2\pi k}{N}\right)
\quad ; \quad E(\theta) = \min_{k} E_k(\theta) 
\end{equation}
so that $ E(\theta) = E(\theta+2\pi)$ while each $E_k(\theta)$ 
is periodic in $2\pi N$. See Fig~\ref{fig_multiv} for $N=10$.

Now, the domain wall tension between different `$k$-vacua' is $O(N)$ 
so as $N\to\infty$ these will all become stable
\cite{EW_theta,shifman_theta,az}
and there are arguments this will happen already at modest
$N$ (=3?). So if we look at $\theta = 0$ in Fig~\ref{fig_multiv},
we see that we have a whole tower of nearly stable vacua that
are above the true vacuum, and which will, at some other 
$\theta$, themselves become the true vacua.

There are some things we have already learned from the lattice.\\
$\bullet$ $T \leq T_c$ : we expect $N$ vacua at any given $\theta$,
and there is some nice lattice evidence for this scenario in 
\cite{pisa_theta}
where these authors show that the periodicity of our usual vacuum
is indeed much greater than the naive $2\pi$, by the simple yet 
effective astuce of calculating high moments of $Q$.\\
$\bullet$ $T > T_c$ : topological fluctuations  disappears roughly 
exponentially in $N$
\cite{blmtuwQ,pisaQ},
as you can see in the right hand plot
\cite{blmtuwQ}
of  Fig.~\ref{fig_multiv}. 
So in the deconfined phase there need be no interlaced vacua -- 
just naive $2\pi$ periodicity in $\theta$ with exponentially
small $E(\theta)$ variation.

So as $N\to\infty$ there are $N$  stable vacua at $\theta=0$.
Now at large $N$ and at  $\theta=0$, the lowest of these
new vacua are close to their minima, and there we can use a quadratic 
approximation for $E_k(\theta)$, giving :
\begin{equation}
E_k(\theta=0) = E_0(\theta=0)+\frac{1}{2} \chi_t (2\pi k)^2
\qquad : \quad k\ll N
\end{equation}
where $\chi_t = \langle Q^2 \rangle/V$ is the topological susceptibility.
Thus the gap to the first of these vacua is
\begin{equation}
E_{k=1} - E_0  = 2\pi^2\chi_t \sim (360 MeV)^4 
\end{equation}
in contrast to the $ E_0 \sim -O(N^2)$ vacuum energy. This is a
modest gap that will be much smaller, at even moderate $N$, than,
for example, the latent heat of the confinement-deconfinement 
transition, $L_h \sim 0.77 N^2 T^4_c$
\cite{blmtuwT}.

These near-stable vacua should appear as
quasistable in computer time when using local Monte Carlo updates 
and we should be able to find them! 
How might we identify such $k$-vacua? 

Firstly, in a $k\neq 0$ vacuum $\langle Q \rangle \neq 0$
since these vacua are not $CP$-invariant, $ k \stackrel{CP}{\to} N-k $,
although they come in degenerate pairs and will therefore mix
-- although there is a large barrier. This signal will be 
tricky to make use of
at large $N$, where transitions between sectors of different
topological charge are suppressed exponentially in $N$ (the
famous $\propto \exp\{-8\pi N/(g^2(\rho) N)\}$ factor for small
instantons) because for $Q$
to change, an instanton has to become small before it shrinks out of
a hypercube (or the reverse). So unless we work at a very coarse $a$,
sequences of configurations will get locked into fixed $Q$ so
that $\langle Q \rangle$ becomes incalculable.

Secondly, we expect the string tension to decrease with increasing 
$k$, since the vacuum energy (the `bag constant') increases.
This might go as
\cite{az}
$ \sigma(k) \simeq \sigma(k=0)\cos\{\frac{2\pi k}{N}\}  $.
(Suggesting that the upper half of vacua,
$N/4 \leq k \leq 3N/4$, are unstable.) To use this we need
enough metastability to be able to acquire high statistics
in this false ($k$-)vacuum.

Thirdly, since the vacuum energy increases with
$k$, $E(k)=E(k=0)+O(k^2)$, perhaps such a state deconfines 
at a lower temperature: $T_c(k\neq 0) < T_c$?

So how do you produce such vacua? One possibility is to `quench' 
in $\beta$ across the strong-weak coupling bulk transition.
Here the latent heat is $O(N^2)$ and one might hope to fall
into one of these confining $k$-vacua more or less at random.
I should add that I tried this a few years ago but could not make 
it work. (It might be because the strong coupling phase has a 
small correlation length, so the configuration after quenching 
is full of bubbles, and the usual vacuum bubble then grows and 
takes over.) Another, probably more promising possibility is 
to `quench' in $\beta$ from  $T>T_c$ to low $T$.

Note that one of the fascinating aspects of these vacua is
that they are almost certainly the pure gauge counterparts
of the degenerate vacua in the confining phase of SU($N$)
${\cal N}=1$ SUSY. To be specific, start in the latter theory,
where one has a set of massless adjoint fermions (gluinos).
There is an anomaly and a spontaneous symmetry breaking that
leads to a gluino condensate
\cite{N1susy}
whose phase is an element of the center and labels the
degenerate vacua. Now make the gluino mass finite, $m_{qA}\neq 0$,
thus explictly breaking the degeneracy. Now take  $m_{qA}\to\infty$,
so that the gluinos decouple and we are left with the pure gauge theory.
The idea is that as we do this
\cite{shifman_theta}
the originally degenerate vacua become the $k$-vacua we have been discussing
here. It would be of great interest, to a much wider theory community
than the one we usually address with lattice QCD, if we were 
to explicitly demonstrate such a scenario in a lattice calculation.

\section{Flux tubes as strings}

The idea that the strong interactions might be formulated
as a string theory, is older than QCD. Indeed string theory
began with the Veneziano model (late 60's) and for the first few 
years was being developed as a theory of the strong interactions.
The fact that at large $N$ diagrams can be naturally mapped 
to a sequence of manifolds that look like a perturbative 
expansion in string theory, and that at strong 't Hooft coupling
the vertices become dense over these manifolds, suggests
('t Hooft, mid-70's) that at large-$N$, and particularly at strong 
coupling, the gauge theory might be a string theory. This is
a limit that has  recently been resuscitated in the gauge-gravity
duality which provides a precise string description for certain 
large-$N$ gauge theories at strong coupling (Maldacena, late 90's). 
And finally,
the string-like character of the confining flux tube, and
the fact that Wilson loops are natural variables for a confining
gauge theory, also motivated the construction of effective
string actions (Polyakov, late 70's).

The natural starting point for building up a stringy description
is to investigate the properties of confining flux tubes. This has 
been an active research area from the earliest days of lattice 
calculations. As a result of this, it is now fairly clear that 
long flux tubes can be described by an effective string theory 
\cite{mlpw_ps}
that is
in the universality class of the free bosonic string theory, i.e.
Nambu-Goto in flat space-time. In recent years there has been
a great deal of work in both $D=3+1$ and $D=2+1$ (which are, for 
these purposes, equally interesting). For example, comparing
Wilson loops to Nambu-Goto (Caselle, Gliozzi, ...), SU(3) string
spectrum (Kuti, ...), potentials in $D=2+1$ both for SU(2) (Hari Dass,
Majumdar, ...) and SU(5) (Meyer). Here I will briefly describe
a current programme in this area that I have been involved in, with 
Barak Bringoltz and Andreas Athenodorou. It is for SU($N$) gauge
theories in $D=2+1$ and it is currently being extended to  $D=3+1$.

If we just want to consider the properties of a confining flux tube,
it is convenient to have a set-up without sources. So suppose we
want a flux tube of length $l$ that closes on itself. 
To ensure its stability, make it wind around a 
spatial torus chosen to have size $l$. The correlators will involve
operators that are variants of smeared/blocked Polyakov loops
that wind around the spatial (not temporal!) torus. Other lattice
dimensions are made effectively $\infty$. There is a phase transition 
at $l_c\equiv 1/T_c \simeq 1.1/\surd\sigma$ such that for $l < l_c$
we have no confining flux tube. So we can study the spectrum of
such flux tubes for all $l > l_c$ and compare them to simple string 
expectations.

\begin	{figure}[h!]
\begin	{center}
\leavevmode
\centerline{\scalebox{0.70}{\epsfig{figure=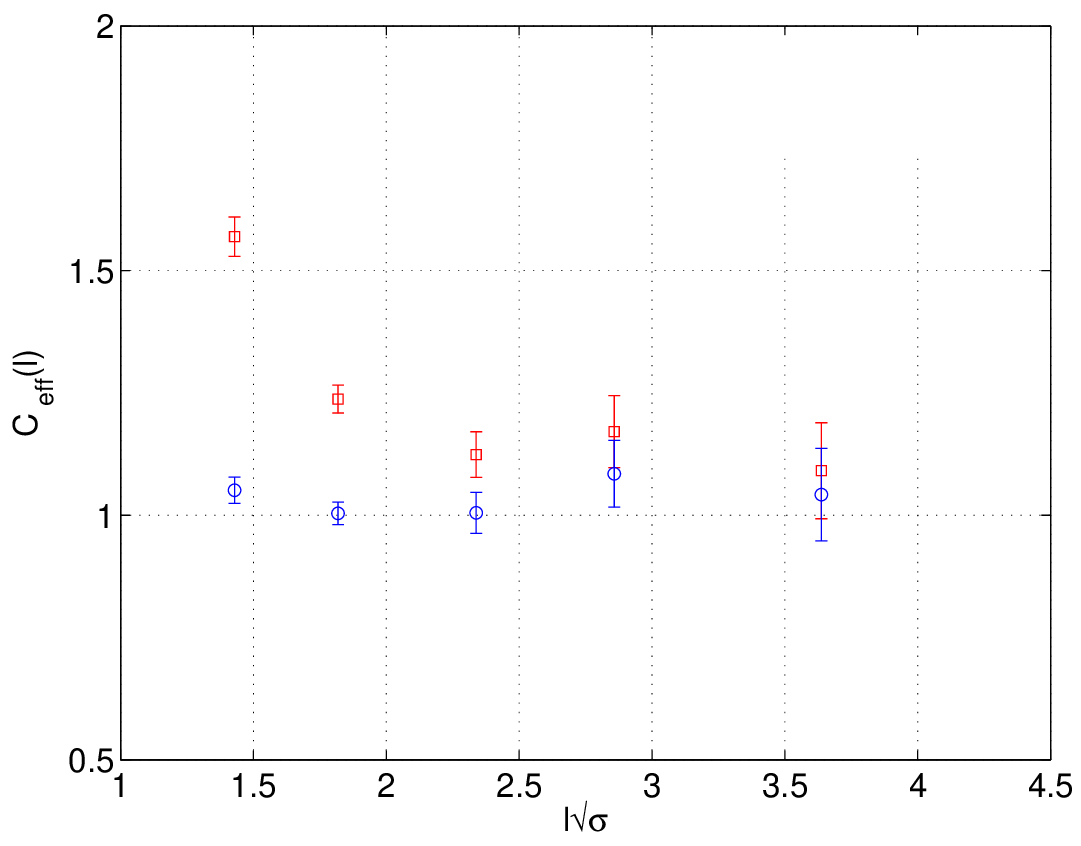, angle=360, width=10cm} } \hspace{-5mm}  {\input	{plot_Ceff12su2b}}}
\end	{center}
\vskip -0.5cm
\caption{Effective Luscher and Nambu-Goto coefficients for SU(5) (left)
and SU(2) (right).}
\label{fig_ceffd3}
\end 	{figure}

We start with the ground state of the flux tube
\cite{es_k1,gs_k}.
In Fig.~\ref{fig_ceffd3}
I show the effective Luscher coefficent $c_{eff}$ defined in
eqn(\ref{eqn_luscher}). On the left SU(5), open squares; on the
right SU(2) and open circles. The approach to the free bosonic
string value, $c_{eff}=1$, looks unambiguous, and very similar
for SU(2) and SU(5). The other sets of points shown are obtained
by modifying the ground state energy of the Nambu-Goto string, i.e.
\begin{equation}
E_0(l)
=
\sigma l 
\left(1 -  c_{eff}(l)\frac{\pi}{3}\frac{D-2}{\sqrt{\sigma}l^2}
\right)^{\frac{1}{2}}.
\label{eqn_NGgs} 
\end{equation} 
where again  $c_{eff}(l)=1$ would be the Nambu-Goto value. 
As we see $c_{eff} =1$ for almost all $l$, i.e. the free string
expression describes flux tubes down to $l\surd\sigma \sim1$ where
they are presumably fat blobs, not slim tubes at all! 
Note that when we expand $E_0(l)$ in eqn(\ref{eqn_NGgs}) in powers 
of $1/{\sqrt{\sigma}l^2}$ we get the Luscher correction in
eqn(\ref{eqn_luscher}) as the first correction. So what
we have learned here is that the higher order corrections are
also Nambu-Goto to a very good approximation. In fact we now
know theoretically 
\cite{ml_jd}
that the next term in the expansion is universal and the same
as in Nambu-Goto. However the agreement we are seeing goes
much further than that: if we fit with Nambu-Goto and include
the first non-universal correction we find that its coefficient
is unnaturally small, $C < O(0.1)$, when expressed in natural units.

\begin	{figure}[h!]
\begin	{center}
\leavevmode
\input	{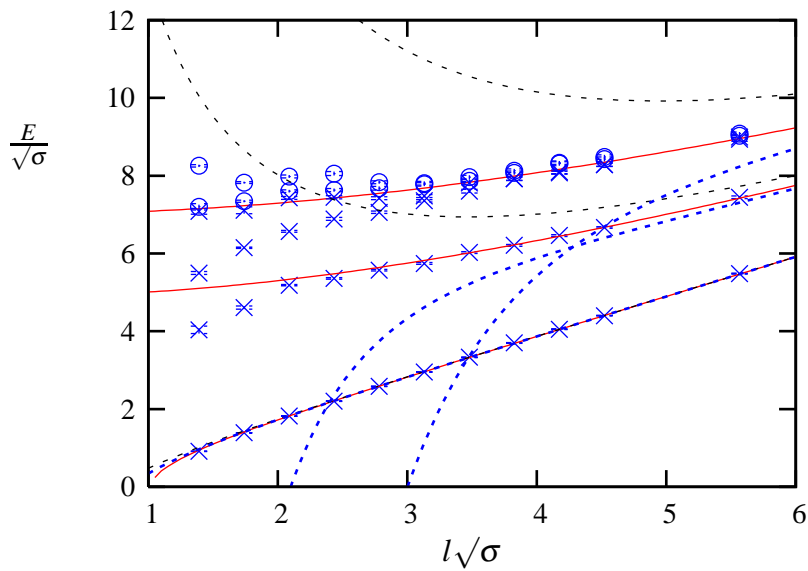}
\end	{center}
\vskip -0.5cm
\caption{Flux loop spectrum in SU(3): solid lines Nambu-Goto,
dashed lines see text. }
\label{fig_esk1}
\end 	{figure}

To show how much further this goes, look at the low-lying
flux loop spectrum in Fig.~\ref{fig_esk1} 
\cite{es_k1}.
(For zero momentum along the flux tube.) The solid lines are 
the predictions of Nambu-Goto. Essentially the excitations 
correspond to free massless `phonons' travelling along
the flux loop, with momenta that are quantised by the periodicity 
of the string. The ground and first excited state are predicted
to have parity $P=+$ and the second excited level has two $P=+$ 
and two $P=-$ states that are degenerate. The numerical results
in  Fig.~\ref{fig_esk1} reproduce precisely this pattern
of degeneracies and quantum numbers.  Note that the only parameter 
in Nambu-Goto is the string tension which is obtained
from the ground state fit. The excited states are thus  predictions 
with no free parameter at all. I invite you to be surprised that
a flux tube of length $l\surd\sigma \sim 2$, surely a blob, 
has a first excited state that is precisely given by the
oscillation of a thin string.

In Fig.~\ref{fig_esk1} the long-dashed lines are what you get with
just the first (Luscher) correction, and the short-dashed lines
include the next universal correction. These are clearly no use for
excited states in this region of $l$. The reason is that for
most of this range, the expansion in $1/\sqrt{\sigma} l^2$
is divergent for these excited states. (As is the expansion
of the Nambu-Goto expression.)

So the message is that the starting approximation one should use is 
the Nambu-Goto spectrum, rather than some large-$l$ truncation thereof.
It is interesting that $k=2$ flux tubes, which are bound states
of two $k=1$ flux tubes, do show $O(1)$ corrections to Nambu-Goto,
as one would expect in a long-distance effective string description
\cite{gs_k,es_k}.
This contrast highlights the significance of what one sees here
for the fundamental flux tube. This suggests that somehow 
even short blobby fundamental flux tubes know that
they are really strings. This is not natural in an arbitrary
`Nielsen-Olesen vortex' type of picture. But it is what happens
in gauge-gravity duals.

It is now of great interest to determine the qualitative features
of the inter-phonon interactions, and to try and find non-string
(`breathing'?) modes that should be there, even in a gauge-gravity
dual picture, and which might give hints as to the appropriate gravity
setup. Returning to other methods, Wilson loop calculations are
effectively a natural transform of the eigenspectrum, and might reveal
features that are not apparent when looking at the low-lying spectrum.
And potential calculations with static sources, if performed
at larger $N$, might reveal something very interesting about the
way the $N=\infty$ theory matches together the IF confining physics
with the UV asymptotically free physics.

\section{Conclusions}

Large-$N$ is at the intersection of much interesting theory and
phenomenology. I have, in my talk, pointed to some of the 
questions that are doable and interesting -- but there is more
in the large number of topics that I have omitted. 
I would re-emphasise that many precise and informative calculations are 
possible with resources readily accessible to almost anyone -- and
really definitive calculations to the many lattice groups with 
Teraflop resources.

\end{document}